\documentstyle[12pt]{article}
\newcommand{\II}{{\cal I}}
\newcommand{\wt}{\widetilde}

\newcommand{\be}{\begin{equation}}
\newcommand{\ee}{\end{equation}}
\newcommand{\ben}{\begin{eqnarray}\displaystyle}
\newcommand{\een}{\end{eqnarray}}
\newcommand{\refb}[1]{(\ref{#1})}
\newcommand{\p}{\partial}
\newcommand{\sectiono}[1]{\section{#1}\setcounter{equation}{0}}

\begin{document}

{}~ \hfill\vbox{\hbox{hep-th/9605150}\hbox{MRI-PHY/96-14}}\break

\vskip 3.5cm

\centerline{\large \bf $F$-theory and Orientifolds}

\vspace*{6.0ex}

\centerline{\large \rm Ashoke Sen\footnote{On leave of absence from 
Tata Institute of Fundamental Research, Homi Bhabha Road, 
Bombay 400005, INDIA}
\footnote{E-mail: sen@mri.ernet.in, sen@theory.tifr.res.in}}

\vspace*{1.5ex}

\centerline{\large \it Mehta Research Institute of Mathematics}
 \centerline{\large \it and Mathematical Physics}

\centerline{\large \it 10 Kasturba Gandhi Marg, 
Allahabad 211002, INDIA}

\vspace*{4.5ex}

\centerline {\bf Abstract}

By analyzing $F$-theory on $K3$ near the orbifold limit of $K3$ we
establish the equivalence between $F$-theory on $K3$ and an orientifold
of type IIB on $T^2$, which in turn, is related by a T-duality
transformation to type I theory on $T^2$. By analyzing the 
$F$-theory
background away from the orbifold limit, we show that 
non-perturbative effects in the orientifold theory splits an
orientifold plane into two planes, with non-trivial SL(2,Z)
monodromy around each of them. The mathematical description of
this phenomenon is identical to the Seiberg-Witten result for
N=2 supersymmetric $SU(2)$ gauge theory with four quark flavors.
Points of enhanced gauge symmetry in the orientifold / $F$-theory
are in one to one correspondence with the points of enhanced
global symmetry in the Seiberg-Witten theory.

\vfill \eject

\baselineskip=18pt

\sectiono{Introduction and Summary}

Besides establishing connections between apparently different
string theories, recent developments in string theory have also 
provided us with some novel ways of compactifying string 
theory. One such procedure, now known as 
$F$-theory\cite{VAFA,VMOR,VMORT,SEIWIT,WITO,FERR,ASGR}, involves
type IIB string compactification where the dilaton and the
scalar from the Ramond-Ramond (RR) sector (which we shall refer
to as the axion) are not constant, but
vary on the internal manifold. Given a manifold $M$ that admits
elliptic fibration, {\it i.e.} has the structure of a fiber
bundle whose fiber is a two dimensional
torus and base is some manifold $B$, one defines $F$-theory on
$M$ as the type IIB theory on $B$, with the axion-dilaton
modulus of the type IIB theory being set equal to the complex
structure modulus
of the fiber. Since in general this modulus varies as we move
on $B$, the axion-dilaton modulus of the type IIB theory will
vary as we move on $B$. In particular as we travel along
non-trivial closed cycles on $B$ the fiber can undergo
non-trivial SL(2,Z) transformation. This would imply that
the axion-dilaton modulus of the type IIB theory will undergo
non-trivial SL(2,Z) transformation as we move along closed
cycles of $B$.

Many of the $F$-theory compactifications have been conjectured to
be equivalent to more conventional compactifications of heterotic
string theory. In particular, $F$-theory on an elliptically
fibered $K3$ surface has been conjectured to be dual to heterotic
string theory on a two dimensional torus\cite{VAFA}. 
One of the purposes
of this paper is to establish this equivalence. 
Like many of the dualities in string theory and M-theory which
can be established by working at the orbifold limit of
the compact manifolds\cite{SONE}, the duality
between $F$-theory on $K3$ and heterotic string theory on $T^2$
is established by going to a special point in the $K3$ moduli
space where it can be identified to a $Z_2$ orbifold of a four
torus. We show that in this limit the $F$-theory 
background reduces to a conventional background where the 
axion-dilaton modulus remains constant as we move in the internal
space. More specifically this background can be identified to
that of an orientifold\cite{SAGN,HOR,ORIENT} 
of type IIB theory which can be analyzed by the
conventional conformal field theory techniques and is in fact
related by $T$-duality to type I string theory on 
$T^2$.\footnote{We have been informed by C. Vafa that 
he has independently made a similar observation.} Thus
at this special point in the moduli space $F$-theory on $K3$ is
identical to type I theory on $T^2$, which, in turn has been
conjectured to be equivalent to heterotic string theory on 
$T^2$\cite{WITTEN,DABH,HULL,POLWIT}. 
Once we have established the equivalence between the
heterotic string theory on $T^2$ and $F$-theory on $K3$ at a special
point in the moduli space, we can argue that the equivalence
must hold at all points in the moduli space since we can deform
both theories away from this specific point by switching on
appropriate background fields. 

We also explicitly study deformations of
the $F$-theory, as well as the orientifold theory, away from this
special point in the moduli space. 
The moduli space of the
orientifold theory is characterized by the vacuum expectation
value of the Higgs
field in the adjoint representation of the gauge group,  or
equivalently, locations of the sixteen seven-branes on the internal
two dimensional manifold. On the other hand the $F$-theory moduli
space is characterized by the moduli of elliptically fibered
$K3$ surfaces. Both theories are described by background field
configurations which consist of dilaton-axion modulus with
non-trivial dependence on the coordinates of the
internal two dimensional manifold. Explicit comparison of the
two sets of field configuration reveals that they are identical
in the two theories at weak coupling, but differ for finite 
coupling.
This difference is non-perturbative in the coupling constant of the
orientifold theory. 
If we focus on the physics near one of the four
orientifold planes, then the mathematical description of the field
configuration turns out to be
identical to the Seiberg-Witten solution of the N=2 supersymmetric
$SU(2)$ gauge theory with four quark flavors\cite{SEIWITTWO}. 
In the analysis of ref.\cite{SEIWITTWO} the
moduli space of N=2 supersymmetric $SU(2)$ gauge theory
was characterized by a gauge invariant quantity $u$,
and the complex `coupling constant' $\tau$ varies
as we move in the $u$ plane. The $F$-theory background is identical
to this configuration, with $u$ labelling the coordinate of the 
base $B$ and $\tau$ denoting the 
axion-dilaton modulus. On the other hand the orientifold 
background corresponds to the classical limit of this 
configuration. As is well known from the analysis of 
ref.\cite{SEIWITTWO}, the classical limit of
this background is singular as $Im(\tau)$
becomes negative in some regions. Thus we expect the $F$-theory
background to describe the quantum corrected version of the
orientifold background.

The $SU(2)$ gauge theory with four quark flavors
is characterized by five
parameters, $-$ the asymptotic value
of $\tau$, and the four quark masses
in the Yang-Mills theory. In the orientifold theory, these four
mass parameters denote  the locations of the four seven
branes around an orientifold plane. (These
in turn can be related to the vacuum expectation value of
a scalar field belonging to the adjoint representation of
the gauge group $SO(8)$.)
As was discussed
in ref.\cite{SEIWITTWO}, at special points in the space
of the parameters $m_i$ the N=2 supersymmetric
Yang-Mills theory has enhanced
global symmetry group $G$. It turns out that precisely at these
special points the orientifold / $F$-theory develops enhanced 
gauge symmetry $G$. 

It was also noted in ref.\cite{SEIWITTWO}
that the SL(2,Z) action on $\tau$ has to be accompanied by 
a triality action on the representations of the global
symmetry group $SO(8)$
which transform the parameters $m_i$ in a non-trivial manner.
Thus we would expect that in the orientifold / $F$-theory, the
SL(2,Z) action on the axion-dilaton modulus will have to be
accompanied by triality action on the representations of the
gauge group $SO(8)$, and in particular on the Higgs vacuum
expectation values (locations of seven branes) represented by
the parameters $m_i$.
We explicitly verify this triality action of SL(2,Z) in the
dual heterotic string theory on $T^2$ where SL(2,Z) is part of the
T-duality group of the theory.

Finally, by comparing the masses of BPS states in the $F$-theory
and in the orientifold theory, we show that in $F$-theory the masses
of BPS states can be expressed in terms of period integrals of the
holomorphic two form on the (complex) surface on which $F$-theory
is compactified. Whenever one or more of the period integrals
vanish, the corresponding BPS state(s) become massless, and at the
same time the surface becomes singular. This is a reflection of the
relationship between singularities of the surface and the appearance
of enhanced gauge symmetries in the corresponding $F$-theory 
compactification.

The paper is organized as follows. Section 2 is devoted to
studing the $F$-theory on $K3$ at the orbifold limit
of $K3$ and comparing the background field configuration 
describing this $F$-theory compactification with the background of an
orientifold of type IIB theory on $T^2$. In section 3 we study
deformation of both, the $F$-theory and the orientifold
backgrounds away from this special point in the moduli space.

\sectiono{$F$-theory on $K3$ in the Orbifold Limit}

Let us begin with the following elliptically fibered $K3$
surface
\be \label{e1}
y^2=x^3+f(z) x + g(z)
\ee
where $x,y,z\in CP^1$, $f(z)$ is a polynomial of degree eight,
and $g(z)$ is a polynomial of degree twelve in $z$. This describes 
a torus for each point
on $CP^1$ labelled by the coordinate $z$. The modular
parameter $\tau(z)$ of the torus is determined in terms of
the ratio $f^3/g^2$ through the relation
\be \label{e2}
j(\tau(z)) = {4\cdot (24 f)^3\over 27g^2+ 4f^3}\, ,
\ee
where
\be \label{edefj}
j(\tau)={\big(\theta_1^8(\tau)+\theta_2^8(\tau)
+\theta_3^8(\tau)\big)^3\over \eta(\tau)^{24}}\, .
\ee
By definition, compactification of $F$-theory on this
particular $K3$ corresponds to compactification of type
IIB theory on $CP^1$ labelled by $z$, with
\be \label{e2a}
a(z) + i e^{-\Phi(z)/2}= \tau(z)\, .
\ee
Here $a$ denotes the RR scalar field and 
$\Phi$ denotes the dilaton field. Physically such a 
background corresponds to a configuration of twenty four 
seven branes of the type IIB theory transverse to $CP^1$ and
situated at the zeroes of
\be \label{e3}
\Delta\equiv 4f^3 + 27 g^2\, .
\ee
In the generic case, the twenty four zeroes of $\Delta$
are distinct from each other and neither $f$ nor $g$
vanishes at the zeroes of $\Delta$. If $z_i$ denotes
such a zero of $\Delta$ then from eq.\refb{e2} and \refb{e3}
we see that near such a point
\be \label{e4}
j(\tau(z))\sim {1\over z-z_i}\, .
\ee
This gives
\be \label{e4a}
\tau(z) \simeq {1\over 2\pi i} \ln(z-z_i)\, ,
\ee
up to SL(2,Z) transformation.

We shall consider a special point in the moduli space of
this compactification where $\tau(z)$ is independent of
$z$. From eq.\refb{e2} we see that this requires
\be \label{e5}
f^3/g^2 = \hbox{constant}\, .
\ee
Since $g$ and $f$ are polynomials in $z$ of order twelve
and eight respectively, the solution to eq.\refb{e5} is
given by
\be \label{e6}
g=\phi^3, \qquad \qquad f = \alpha \phi^2\, ,
\ee
where $\alpha$ is a constant and $\phi$ is a polynomial
in $z$ of degree four. By a rescaling of $y$ and $x$
we can
set the coefficient of $z^4$ in $\phi$ to be one. Thus
$\phi$ has the form\footnote{A special case of this where
$z_1=z_2=0$ and $z_3=z_4=\infty$ has been discussed in
\cite{VMORT}.}
\be \label{e7}
\phi=\prod_{i=1}^4 (z-z_i)\, 
\ee
where $z_i$ are constants. From eqs.\refb{e2},
\refb{e3}, \refb{e6} and \refb{e7} we get
\be \label{e8}
\Delta=(4\alpha^3+27)\prod_{i=1}^4 (z-z_i)^6\, ,
\ee
and
\be \label{e8a}
j(\tau)= {4\cdot (24\alpha)^3\over 27+ 4\alpha^3}\, .
\ee
Thus this particular compactification corresponds to
a configuration where the twenty four seven-branes are
grouped into four sets of six coincident seven-branes,
situated at the points $z_1,\cdots z_4$. $\tau$ is
constant over $CP^1$; however there is an SL(2,Z)
monodromy 
\be \label{e10}
\pmatrix{-1 & \cr & -1}\, ,
\ee
around each of the points $z_i$. This can be seen by
noting that as $z$ moves once around the point $z_i$,
$y$ changes sign. This corresponds to the hyperelliptic
involution of the torus, represented by the SL(2,Z) 
matrix given in eq.\refb{e10}.

The metric on the base can be read out from the formulae
derived in ref.\cite{COSMIC}. Up to an overall 
multiplicative constant, it is given by
\be \label{emetric}
ds^2={dzd\bar z\over \prod_i (z-z_i)^{1/2} (\bar z - 
\bar z_i)^{1/2}}\, .
\ee
Thus there is a deficit angle of $\pi$ at each of the 
points $z_i$. Thus the base has the geometry of $T^2/\II_2$,
where $\II_2$ acts on the torus by inverting the sign of both
the coordinates of the torus. The modular parameter $\lambda$
of the torus is determined in terms of the cross ratio
\be \label{e9}
{(z_1-z_2)(z_3-z_4)\over (z_1-z_3)(z_2-z_4)}\, ,
\ee
which is invariant under an SL(2,C) transformation of the base
$CP^1$.

This background can now be given an orientifold interpretation
as follows. By studying the action of various symmetry
transformations on the massless fields of the theory it is
easy to verify that the SL(2,Z) transformation \refb{e10}
can be identified to the transformation 
$(-1)^{F_L}\cdot\Omega$ of the type IIB theory, where
$\Omega$ denotes the orientation reversal transformation
(exchange of left and right moving modes on the world
sheet) and $(-1)^{F_L}$ changes the sign of all the Ramond
sector states on the left. Thus we have type IIB 
compactification on $T^2/\II_2$ such that as we go once around
each fixed point on $T^2/\II_2$ the theory comes back to itself
transformed by the symmetry $(-1)^{F_L}\cdot\Omega$. In other
words,
the theory can be identified to type IIB on $T^2$, modded
out by the $Z_2$ transformation 
$(-1)^{F_L}\cdot\Omega\cdot\II_2$. This is an orientifold\cite{ORIENT},
and, as we shall see, is related to type I theory on $T^2$ by a
$T$-duality transformation. We shall denote this
theory as type IIB on $T^2/(-1)^{F_L}\cdot\Omega\cdot\II_2$.
In this case, each of the four orientifold planes carry $-4$
units of seven-brane charge, which need to be neutralized
by putting sixteen seven branes transverse to 
$T^2/\II_2$\cite{ORIENT}.
At a generic point in the moduli space, where the seven-branes
are located at arbitrary positions on $T^2/\II_2$, the
seven brane charges are not neutralized locally and as a
result $\tau(z)$ varies on $T^2/\II_2$. On the other hand,
in order that the
field configuration matches with the one obtained from 
$F$-theory, $\tau(z)$ must be constant on $T^2/\II_2$.
Thus the seven-brane charges must be neutralized pointwise
on $T^2/\II_2$. This happens if the sixteen
seven branes are grouped into four sets of four
coincident seven-branes, and these four sets are placed
at the four orientifold planes. This would give a field
configuration identical to the one obtained from the
$F$-theory configuration.

This establishes the equivalence between $F$-theory
compactification on $K3$ and an orientifold of type IIB
theory at a special point of the moduli space. By making
an $R\to (1/R)$ duality transformation on both the circles
of $T^2$ we can map the $Z_2$ transformation 
$(-1)^{F_L}\cdot\Omega\cdot\II_2$ to the transformation
$\Omega$.\footnote{This fact has been independently observed
in ref.\cite{DPTWO}.} Since modding out the type IIB theory by 
$\Omega$ produces type I theory, we see that the orientifold 
is related by $T$-duality to type I theory on $T^2$. This in
turn has been conjectured to be equivalent to heterotic string theory
on $T^2$\cite{WITTEN,DABH,HULL,POLWIT}. 
Thus through this chain of arguments we have been
able to establish the conjectured duality between $F$-theory
compactification on $K3$ and heterotic string theory on 
$T^2$.
Although this duality was established only at one point in the
moduli space, we can deform both theories away from this
special point by switching on appropriate background fields,
and hence the duality must hold at all points in the moduli
space.

We can also study the enhanced non-abelian gauge symmetries
at this special point in the moduli space, both from the
orientifold view point as well as the $F$-theory viewpoint.
{}From the orientifold viewpoint we have an $SO(8)$ 
gauge symmetry
associated with each orientifold plane, since four seven-branes
and their images meet there. Thus we get an $\big(SO(8)\big)^4$ 
non-abelian gauge symmetry at this special point. On the
other hand, in order to study the enhancement of gauge
symmetry from the $F$-theory viewpoint, we need to study
what kind of singularities appear at the points $z_i$
($1\le i\le 4$) at this special point in the moduli
space. To see this note that after a suitable rescaling
of the various coordinates, eq.\refb{e1} near the point
$z=z_1$ takes the form:
\be \label{e11}
\wt y^2 \simeq \wt x^3 + \alpha \wt x \wt z^2 + \wt z^3\, ,
\ee
where,
\be \label{esing2}
\wt y = y \prod_{i=2}^4 (z_1 - z_i)^{3/2}, \qquad \qquad
\wt x = x \prod_{i=2}^4 (z_1 - z_i), \qquad \qquad
\wt z = (z-z_1)\, .
\ee
This corresponds to a $D_4$ type singularity of the $K3$
surface and hence corresponds to an enhanced $SO(8)$ gauge
symmetry. Since there are four such singular points, we get
a net non-abelian gauge group $\big(SO(8)\big)^4$, in 
agreement with 
the answer obtained from the orientifold analysis.

\sectiono{Deforming away from the Orbifold Limit of $K3$}

In this section we shall discuss deforming both the $F$-theory
and the orientifold theory away from the special 
point in the moduli
space considered in the previous section and compare the results.
In the $F$-theory such a deformation would 
correspond to splitting
the six coincident zeroes of $\Delta$ away from each other. On the
other hand, for the orientifold, this corresponds to moving the 
four coincident seven-branes away from the orientifold plane.
In order to learn the physics of the situation we can focus our
attention on one of the four orientifold planes. Equivalently,
instead of studying $F$ theory on the orbifold $T^4/\II_4$
(where $\II_4$ denotes changing the sign of all the four
coordinates on $T^4$) and the type IIB on $T^2/(-1)^{F_L}\cdot
\Omega\cdot\II_2$, we consider $F$-theory on $T^2\times R^2/\II_4$
and type IIB on $R^2/(-1)^{F_L}\cdot\Omega\cdot\II_2$. Let $z$
denote the complex coordinate on $R^2/\II_2$, 
and $z=0$ denote the fixed
point, so that at the special
point in the moduli space
condsidered in the previous section the metric takes the form:
\be \label{e11a}
ds^2=dzd\bar z/z^{1/2}\bar z^{1/2}\, .
\ee
Also let $\tau_0$ be the constant
value of $\tau$ away from the singular
point. We 
shall study the deformation of this background keeping
the asymptotic metric fixed to be of the form \refb{e11a} and
asymptotic
$\tau$ fixed at $\tau_0$. The collective
coordinates of this theory describe an 
N=1 supersymmetric $SO(8)$
gauge theory in eight dimensions. The moduli space of this
theory is characterized by the vacuum expectation value of the
complex scalar field $\phi$
belonging to the adjoint representation
of the gauge group. 
At a generic point in the moduli space this
vacuum expectation value takes the form:
\be \label{evev1}
\langle \phi\rangle =
\pmatrix{i\sigma_2 c_1 &&&\cr & i\sigma_2 c_2 && \cr
&& i\sigma_2 c_3 & \cr &&& i\sigma_2 c_4\cr}, \qquad \qquad
\sigma_2 = \pmatrix{0 & -i \cr i & 0}\, ,
\ee
where $c_i$ are complex parameters.

First let us describe this deformation in the orientifold
theory. In this case the deformed configuration will
correspond to an orientifold plane carrying $-4$ units of seven-brane
charge (which we shall take to be fixed at $z=0$)
and four seven branes at arbitrary 
coordinates $z_i$. Note that if we use the freedom of rescaling
$z$ we can eliminate one of the $z_i$'s from the set of
independent parameters, but we do not use this freedom as this 
will change the
asymptotic form of the metric. These four complex parameters
can be related to the parameters $c_i$ introduced earlier by
working in the cordinate 
\be \label{e17}
w=z^{1/2}\, ,
\ee
which is the natural coordinate on $R^2$ (as opposed to $z$ which
is the natural coordinate on $R^2/\II_2$). In this coordinate
system the orientifold plane carries $-8$ units of seven brane
charge, and there are eight seven branes on $R^2$ distributed
in an $\II_2$ invariant fashion. $\pm c_i$ are the locations of 
these eight seven branes in $R^2$\cite{ORIENT}. Thus we have 
\be \label{ezc}
z_i=c_i^2\, .
\ee

Since the seven-brane charge is
no longer neutralized locally, $\tau$ is no longer a constant 
in the $z$ plane. Naively, we would expect it to be of the form:
\be \label{e12}
\tau(z)= \tau_0 +{1\over 2\pi i} \Big( \sum_{i=1}^4
\ln(z-z_i) - 4\ln z\Big)\, ,
\ee
since there is $+1$ unit of seven brane charge at each $z_i$,
and $-4$ unit of seven brane charge at $z=0$.
This solution is characterized by five parameters, $-$ $\tau_0$ and
the four $z_i$'s. A closer examination of the solution reveals
however that the solution does not make sense everywhere in the
$z$ plane. In particular, close to $z=0$, $Im(\tau)$ becomes
large and negative, violating the bound $Im(\tau)\ge 
0$.\footnote{The situation is similar to the result of 
ref.\cite{POLWIT} in one compact dimension, where the supergravity
solution broke down for a sufficiently large separation between
the orientifold plane and the eight branes. In the present case the
solution breaks down for arbitrarily small but finite separation
between the orientifold plane and the seven branes due to 
stronger short distance divergence in two dimensions.}
Thus we would expect that strong coupling effects will modify
the solution near the $z=0$ point. 
We shall now see that the $F$-theory description of the background
provides us with precisely such a modification. Note however
that due to the high degree of supersymmetry present in the 
problem, we expect the moduli space of the theory to remain
unmodified by quantum corrections, at least locally. Thus
$c_i$ (or equivalently $z_i$) should continue to label the
moduli space of the theory, but the background field
configuration describing the theory will no longer be of the
form \refb{e12}.

{}From the $F$-theory viewpoint, the deformation of $T^2\times 
R^2/\II_4$ away from the orbifold limit is described by
a surface of the form
\be \label{ebb1}
y^2 = x^3 + \wt f(z) x + \wt g(z)\, ,
\ee
where $\wt f$ and $\wt g$ are now
polynomials in $z$ of degree two and three respectively. This
gives a total of seven complex parameters to begin with. Of
this one parameter can be removed by an overall shift of $z$
and another can be removed by a rescaling of $x$ and $y$. This
again leaves us with five complex parameters. In order to relate
these five parameters to those obtained in the orientifold
description it is convenient to choose these five parameters in
a specific manner. 
In the analysis
of N=2 supersymmetric $SU(2)$ gauge theory with
four quark flavors\cite{SEIWITTWO}
Seiberg and Witten were led to a similar surface parametrized
by four quark masses $m_i$ ($1\le i\le 4$) and the complex
coupling constant $\tau_0\equiv (\theta/\pi + 8\pi i/g^2)$. 
The surface was given by
\be \label{ee16}
y^2=W_1W_2W_3 + A(W_1 T_1 (e_2-e_3)+W_2T_2(e_3-e_1)
+W_3T_3(e_1-e_2)) - A^2 N\, ,
\ee
where
\be \label{e16a}
W_i=x-e_i \wt z -e_i^2 R, \qquad \qquad A=(e_1-e_2)(e_2-e_3)(e_3-e_1)
\, ,
\ee
\be \label{ee17}
e_1-e_2=\theta_3^4(\tau_0), \qquad e_3-e_2=\theta_1^4(\tau_0),
\qquad e_1-e_3=\theta_2^4(\tau_0), \qquad e_1+e_2+e_3=0\, ,
\ee
\be \label{ewtz}
\wt z = z -{1\over 2} e_1 R\, ,
\ee
\ben \label{ee18}
R &=& {1\over 2}\sum_i m_i^2\, , \nonumber \\
T_1 &=& {1\over 12} \sum_{i>j} m_i^2 m_j^2 -{1\over 24} 
\sum_i m_i^4\, ,
\nonumber \\
T_2 &=& -{1\over 2} \prod_i m_i -{1\over 24} \sum_{i>j} m_i^2 m_j^2
+{1\over 48} \sum_i m_i^4\, , \nonumber \\
T_3 &=& {1\over 2} \prod_i m_i -{1\over 24} \sum_{i>j} m_i^2 m_j^2
+{1\over 48} \sum_i m_i^4\, , \nonumber \\
N &=& {3\over 16} \sum_{i>j>k} m_i^2 m_j^2 m_k^2 -{1\over 96}
\sum_{i\ne j} m_i^2 m_j^4 +{1\over 96} \sum_i m_i^6\, .
\een
The surface described by 
eq.\refb{ee16} is not exactly of the form
\refb{ebb1} since the coefficient of the $x^2$ term does not vanish,
but this can be removed by an overall $z$ independent shift in the
coordinate $x$. We shall use the same set of parameters $\{m_i\}$ 
and $\tau_0$ to label the $F$-theory background, and
show that we get a consistent map between
the orientifold and the $F$-theory by postulating the 
following simple relation between the parameters $m_i$
labelling the $F$-theory and the 
parameters $c_i$ labelling the orientifold:
\be \label{emc}
m_i=c_i\, ,\qquad \qquad 1\le i\le 4\, .
\ee
This would imply that up to an $SO(8)$ gauge transformation,
the vacuum expectation value of the adjoint scalar $\phi$
is given by
\be \label{evev}
\pmatrix{i\sigma_2 m_1 &&&\cr & i\sigma_2 m_2 && \cr
&& i\sigma_2 m_3 & \cr &&& i\sigma_2 m_4\cr}, \qquad \qquad
\sigma_2 = \pmatrix{0 & -i \cr i & 0}\, .
\ee

In the analysis of Seiberg and Witten the coordinate $z$ on the
base represented the gauge invariant modulus representing the
square of the Higgs expectation value. There are six
singularities in the
$z$ plane, which, in ref\cite{SEIWITTWO}, signalled the
appearance of massless charged particles
at these points in the moduli space. In the weak coupling (large
$Im(\tau_0)$) limit, four of these singularities are located
at $z\simeq m_i^2$ representing points where the four quarks become
massless. The two other singularities are located close to the
origin within a distance of order $\exp(i\pi\tau_0/2)$, and
represent points where massless monopoles and dyons appear in
the spectrum. Thus in the $\tau_0\to i\infty$ limit, these two
singularities coalesce, and we recover the semiclassical picture
where the $SU(2)$ gauge symmetry is restored at the origin and the
$W^\pm$ states become massless. Indeed, for large $Im(\tau_0)$
and $|z|>>|e^{i\pi\tau_0/2}|$ the solution
described in \refb{e16a} can be rewritten as
\be \label{e16}
\tau(z)\simeq\tau_0+{1\over 2\pi i} \Big(\sum_{i=1}^4\ln(z-m_i^2) 
- 4\ln z\Big)\, .
\ee
The coefficient 4 in front of $\ln z$ is due to the fact that the
$W^\pm$ bosons carry twice the electric charge of a quark, and 
the relative $-$ sign between the two terms is due to the fact
that the $W^\pm$ belong to a vector multiplet whose contribution
to the $\beta$-function has sign opposite to that of a
hypermultiplet. This configuation
is identical to the one given in eq.\refb{e12} describing the
orientifold configuration, provided we identify $z_i$ with
$m_i^2$, {\it i.e.} $c_i$ with $m_i$. 
This shows that in the weak coupling limit the 
background field configuration corresponding to an orientifold
is identical to the one described by $F$-theory even away from
the special point in the moduli space described in the last 
section. 
However, the background field configuration describing
the orientifold breaks down close to the orientifold point where
the coupling constant becomes strong. On the other hand the
$F$-theory background makes sense everywhere in the $z$ plane. This
leads us to conclude that the $F$-theory provides the
correct description of the background field configuration of this
theory, and the orientifold background must be
modified by quantum correction so as to coincide with the $F$-theory
background. This would imply in particular that the orientifold
plane $z=0$ is split into two planes due to quantum corrections,
in a manner analogous to the splitting of the $z=0$ point in the
moduli space of N=2 supersymmetric $SU(2)$ gauge theories into two
points. The splitting, being of order $\exp(i\pi\tau_0/2)$,
is non-perturbative in the orientifold coupling constant, and is not
visible in the perturbation theory.

The above analysis shows the equality of the parameters
$m_i$ and $c_i$ in the weak coupling limit where we can
directly compare the orientifold background with the
$F$-theory background. 
We shall now show that eq.\refb{emc} gives the correct map
between the points of enhanced gauge
symmetry in the orientifold and the $F$-theory descriptions
even away from the weak coupling region. In subsection 3.3 we
shall trace the origin of this simple relation
between $m_i$ and $c_i$ to the fact that
$m_i\pm m_j$ are related to the period integrals of the holomorphic
two form on the surface described by eq.\refb{ee16}.

\subsection{Unbroken Gauge Symmetries}
For a Higgs vacuum expectation
value of the form \refb{evev1} the unbroken gauge symmetry is the
subgroup of $SO(8)$ that commutes with this matrix.
In particular, if $n$ of the $c_i$'s are equal and non-zero,
then we recover an $SU(n)$ gauge symmetry, whereas if $n$ of them
are equal and zero, we recover an $SO(2n)$ gauge symmetry.
In the orientifold description these enlarged gauge symmetries
are associated with coincident seven branes\cite{WISU}. 
We shall now show that with the identification of $m_i$ with
$c_i$, we get the same enhanced gauge
symmetries in the $F$-theory at these special points.
In the $F$-theory background described by eq.\refb{ee16}, 
when $n$ of the $m_i$'s are equal and non-zero, $n$
of the zeroes of $\Delta$, representing the point where these
$n$ quarks become massless in the corresponding $SU(2)$ gauge
theory, coincide. Thus $\Delta$ has an $n$th order zero, but 
typically,
neither $f(z)$ nor $g(z)$ vanish  there. According to the table
in ref.\cite{VMORT} this corresponds to an $A_{n-1}$ type
singularity and hence an enhanced $SU(n)$ gauge symmetry.
The case where $n$ of the $m_i$'s vanish is somewhat more
complicated, and we need to carry out the analysis separately
for each $n$. The relevant values of $n$ are 2, 3 and 4. For $n=4$
all $m_i$'s vanish, and the equation of the surface
reduces to the form \refb{e11}.
This corresponds to a $D_4$ type singularity and hence the $F$-theory 
on such a surface has an enhanced $SO(8)$ gauge symmetry. For $n=3$
the singularity structure near the origin is that of N=2
supersymmetric Yang-Mills theory with three massless quarks. 
According to the analysis of Seiberg and Witten, there are two
singularities near the origin. Using the results of 
ref.\cite{SEIWITTWO} one can easily verify that near one of the
singular points $\Delta$ has a fourth order zero with $f$ and $g$
being finite at that point.  According to ref.\cite{VMORT}
this corresponds to an $A_3$ type singularity, and hence an
enhanced $SU(4)\equiv SO(6)$ gauge symmetry. Finally,
for $n=2$, the singularity structure near the origin is that
of an N=2 supersymmetric $SU(2)$ gauge theory with two flavors of
massless quarks. According to the analysis of Seiberg and Witten
this theory again has two singular points near the origin, and
at each of these singular points $\Delta$ has a second order
zero where $f$ and $g$ remain finite. Using the table of
ref.\cite{VMORT} we see that each of these singularities
is of $A_1$ type, and hence we expect an enhanced
$SU(2)\times SU(2)\equiv SO(4)$ gauge symmetry at this point.
This shows that the identification of the parameters $m_i$
with $c_i$ gives us enhanced gauge symmetries 
in the $F$-theory at correct points in the moduli space.

In the parameter space of N=2 supersymmetric $SU(2)$ gauge theories
with four flavors of quarks, there are special points of
enhanced {\it global symmetry}. In particular, when $n$ of the
quark masses are equal but non-zero, we have a global $SU(n)$
symmetry, whereas if $n$ of the quarks are massless, we have an
enhanced global $SO(2n)$ symmetry.
The analysis of the previous paragraph shows that whenever the N=2
supersymmetric $SU(2)$ gauge theory develops an enhanced
global symmetry, the corresponding $F$-theory develops the
same enhanced gauge symmetry. This has a simple interpretation
in view of the identification of $m_i$ with the parameters
$c_i$.  
In the N=2 supersymmetric Yang-Mills theory, the quark
mass matrix is exactly of the form given in \refb{evev}, and
the unbroken global symmetry group at any point in the
parameter space is the subgroup of $SO(8)$ that commutes with
this matrix. But this is precisely the subgroup of $SO(8)$ that
remains unbroken in the corresponding orientifold / $F$-theory 
compactification, since \refb{evev} represents the vacuum 
expectation value of the adjoint Higgs field in this theory. 
Thus we see that there is a one
to one correspondence between the points of enhanced global symmetries
arising in the supersymmetric Yang-Mills theory and the points
of enhanced
gauge symmetries arising in the orientifold / $F$-theory.

\subsection{Triality}
We shall now discuss the observation of ref.\cite{SEIWITTWO}
that SL(2,Z) transformations have a triality action on the
representations of the global symmetry group $SO(8)$. In
particular, the transformation $\tau\to -1/\tau$ acts by exchanging
the vector and the spinor representations of $SO(8)$ and acts on
the mass parameters $m_i$ as
\ben \label{e19}
m_1 &\to& {1\over 2} (m_1 + m_2+m_3+m_4)\, , \nonumber \\
m_2 &\to& {1\over 2} (m_1 + m_2-m_3-m_4)\, , \nonumber \\
m_3 &\to& {1\over 2} (m_1 - m_2+m_3-m_4)\, , \nonumber \\
m_4 &\to& {1\over 2} (m_1 - m_2-m_3+m_4)\, .
\een
This would imply, in particular, that in the specific orientifold
compactification that we are considering, the action $\tau_0\to
-1/\tau_0$ on the asymptotic $\tau$ by itself is not a symmetry
of the theory, but it must also act on the locations $c_i$ ($m_i$)
of the seven branes (or, equivalently,
on the vacuum expectation values of the adjoint Higgs field) 
according to eq.\refb{e19}.

Is there a way to verify this independently? From the point of 
view of the type IIB orientifold, the symmetry $\tau\to -1/\tau$
is a non-perturbative symmetry, and hence the action of this
transformation on the representation of the gauge group
will be difficult to study
explicitly. However, by using the type I $-$ heterotic equivalence
in ten dimensions, or equivalently, the $F$-theory $-$ heterotic
duality in eight dimensions, one can map the non-perturbative
S-duality transformation of the orientifold theory to a
perturbative T-duality transformation in the heterotic string
theory, where the action of this transformation on the
representations of the gauge group can be studied explicitly.
To do this we need to first patch together four copies of the
solution we have been discussing so as to describe type IIB
compactification on $T^2/(-1)^{F_L}\cdot\Omega\cdot\II_2$.
This can easily be done when the distance between the fixed
points is large compared to the distance between any
given fixed point and the four seven-branes around it.
For this we rewrite eq.\refb{ee16} (after a constant shift of $x$
and a suitable rescaling of $x$ and $y$) as
\be \label{enn1}
y^2 = x^3 + x \alpha(\tau_0)\wt f(z, \{m_i\}, \tau_0)
+ \wt g(z, \{m_i\}, \tau_0)\, ,
\ee
where $\alpha(\tau_0)$ is given by eq.\refb{e8a}, and $\wt f$
and $\wt g$ are polynomials in $z$ of degree two and three
respectively, with the coefficients of the leading power in
$z$ set equal to one in both of them. Then the equation of the
full $K3$ surface, obtained by patching together four of these
solutions will be given by
\be \label{enn2}
y^2 = x^3+ x\alpha(\tau_0)\prod_{s=1}^4 \wt f(z -z_s, \{m_i^{(s)}\},
\tau_0) + \prod_{s=1}^4 \wt g(z-z_s, \{m_i^{(s)}\}, \tau_0)\, ,
\ee
where $\wt f$ and $\wt g$ are the same functions that appear in
eq.\refb{enn1}. With the help of an SL(2,C) transformation on $z$,
we can set $(z_1,z_2,z_3)=(0,1,-1)$.
The solution is then characterized by the set of
sixteen $\{m_i^{(s)}\}$ ($1\le i\le 4$, $1\le s\le 4$),
$\tau_0$ and $z_4$.  When all the $m_i$'s
associated with all the four orbifold points vanish,
this $K3$ surface is described by eq.\refb{e1} with 
$g$ and $f$ given by eqs.\refb{e6}, \refb{e7}.
The unbroken symmetry group in this theory is 
$\big(SO(8)\big)^4$.
Since the same $\tau_0$ represents the value of $\tau$ 
away from each of the four fixed points on $T^2/\II_2$, the
$\tau_0\to -1/\tau_0$ transformation must exchange the vector and
the spinor representation of each $SO(8)$. 

We shall now verify this explicitly by mapping this to the dual
heterotic description. For this we regard this as an orientifold,
and make $R\to (1/R)$ T-duality transformation in each of the
compact directions (which we shall denote by the coordinates
$x^8$ and $x^9$) to map this into type I theory on $T^2$. In this
process the RR scalar, that forms the real part of $\tau$, gets
mapped to $B'_{89}$ where $B'$ denotes the rank two anti-symmetric
tensor in the RR sector. Under the type I - heterotic duality, this
gets mapped to $B_{89}$ where $B$ denotes the rank two anti-symmetric
tensor in the heterotic string theory. Thus $\tau$ gets mapped to
the Kahler modulus of the heterotic string theory on $T^2$, and
hence the S-duality group
SL(2,Z) of the orientifold theory gets mapped to the SL(2,Z) T-duality
symmetry of the $SO(32)$ heterotic string theory compactified
on $T^2$. In particular the transformation $\tau\to -1/\tau$ will
correspond to $R\to 1/R$ duality on both the circles in the heterotic
string theory, together with an exchange of the coordinates
$x^8$ and $x^9$. We shall denote this transformation by $\sigma$ and
show that it induces a triality transformation on the
representations on $\big(SO(8)\big)^4\subset SO(32)$.

We need to study the $SO(32)$ heterotic string theory near a point
in the moduli space where the $SO(32)$ gauge group has been broken
down to $\big(SO(8)\big)^4$. 
This corresponds to introducing $SO(32)$ Wilson 
lines $U_8$ and $U_9$ along $x^8$ and $x^9$ given by
\be \label{ewilson}
U_8=\pmatrix{-I_{8} &&& \cr & -I_{8} && \cr && I_8 & \cr &&& I_8}\, , 
\qquad \qquad
U_9=\pmatrix{-I_{8} &&& \cr & I_{8} && \cr && -I_8 & \cr &&& I_8}\, , 
\ee
where $I_n$ denotes $n\times n$ identity matrix. 
We shall analyze this theory using the fermionic description
of the heterotic string theory.
Without the Wilson lines, the $SO(32)$ heterotic
string contains a conformal field theory of 32 free left-moving
Majorana fermions, and this conformal field theory is modded out
by a $Z_2$ transformation that changes the sign of all the
thirty two
fermions\cite{HETEROTIC}. This $Z_2$ modding out is responsible 
for the GSO projection on the left, and the existence of
twisted sector states
belonging to the spinor representation of $SO(32)$. We shall
group the 32 fermions into four groups of eight each and denote
this $Z_2$ transformation as $(----)$ in order to denote that
is acts as $-I_8$ on all four groups of fermions.  Introduction
of the Wilson lines \refb{ewilson} corresponds to further
modding out the theory by a $Z_2\times Z_2$ transformation,
generated by
\be \label{ezt1}
(--++)(x^8\to x^8+\pi)\, ,
\ee
and
\be \label{ezt2}
(-+-+)(x^9\to x^9+\pi)\, .
\ee
We shall use the convention in which the transformation $-I_8$
of $SO(8)$ changes the sign of the vector ($v$) and the conjugate
spinor ($c$)
representations of $SO(8)$ but leaves the spinor ($s$)
representations of $SO(8)$ invariant. Let us now consider an
untwisted sector state that transforms in the vector 
representations of the first two $SO(8)$. We shall denote such
a state by ($v_1v_2$). This state is even under $(--++)$ and
odd under $(-+-+)$. Thus invariance under \refb{ezt1} and
\refb{ezt2} requires that the state carries even unit of
momentum along $x^8$ and odd unit of momentum along $x^9$.
The duality transformation $\sigma$ converts this to a state
carrying odd unit of winding along $x^8$ and even unit of winding
along $x^9$. In other words, this would correspond to a state
in the twisted sector of \refb{ezt1}. Such a state belongs to
the conjugacy class $(s_1s_2)$. 
Similar analysis shows that the
transformation $\sigma$ takes a state in the conjugacy class
$(v_iv_j)$ to the conjugacy class $(s_is_j)$ for all $i,j$. 
(Due to the twisting by the $(----)$ transformation that
commutes with $\sigma$, we do not distinguish between conjugacy
classes $(s_is_j)$ and $(s_ks_l)$ if $(i,j,k,l)$ is a permutation
of the set $(1,2,3,4)$.) This shows that $\sigma$ does exchange
the vector and spinor representations of each of the four $SO(8)$'s
as expected from the $F$-theory description.

{}From the analysis of ref.\cite{SEIWITTWO} we also know that the
transformation $\tau\to \tau+1$ induces an $SO(8)$ parity
transformation. Thus we would expect that
in type IIB on $T^2/(-1)^{F_L}\cdot\Omega
\cdot\II_2$, $\tau\to\tau+1$ must be accompanied by a parity 
transformation in all four $SO(8)$'s. This however corresponds
to the $SO(32)$ gauge transformation
\be \label{eparity}
\pmatrix{I_7 &&&&&&& \cr & -1 &&&&&& \cr && I_7 &&&&& \cr &&& -1 
&&&& \cr &&&& I_7 &&& \cr &&&&& -1 &&\cr &&&&&& I_7 & \cr 
&&&&&&& -1 \cr}\, .
\ee
Since \refb{eparity} by itself is
a symmetry of the theory, in this theory
$\tau\to \tau+1$ is also a symmetry by itself
without being accompanied by any action on the Higgs field. 
As a result we do not expect to see any non-trivial action of
this transformation on the representations of $\big(SO(8)\big)^4$
in the dual heterotic string theory.

\subsection{BPS States} At a generic point in the
moduli space, type IIB on $R^2/(-1)^{F_L}\cdot\Omega\cdot\II_2$
has BPS states representing the $SO(8)$ gauge bosons and
their superpartners that have become massive due to 
Higgs vacuum expectation value of the form \refb{evev1}. 
In the orientifold
description these correspond to elementary strings
starting on one of the four seven branes and ending on
another seven brane. Due to the presence of the orientifold
plane there are two topologically distinct ways an open string
can stretch between two seven branes. We shall call one of them
the direct route, and the other one, differing from the direct
route by one unit of winding around the orientifold plane, the
indirect route.
In the natural coordinate system $w$ on $R^2$ defined in 
\refb{e17} the direct route is a straight line joining the
points $m_i$ and $m_j$ and the indirect route is a straight
line joining the points $m_i$ and $-m_j$.
With suitable convention for the sign of $\{m_i\}$,
the mass of a BPS state
represented by an open string stretched between the
$i$th seven brane and the $j$th seven-brane along the direct
route is given by
$m_i-m_j$. On the other hand the BPS state corresponding to
an open string stretched between the $i$th and the $j$th
seven brane along the indirect route has a mass given by $m_i+m_j$.

{}From our previous discussion about the relationship 
between the orientifold and the $F$-theory background,
it is clear how to represent these BPS states in the
corresponding $F$-theory description. The description
is in fact identical to the one given in the orientifold
theory, except that winding around the orientifold
plane will correspond to winding around the two 
seven-branes (which we shall label as the fifth and the
sixth seven brane) into which the orientifold plane
splits. We would now like to ask if the masses of these
BPS states can be expressed in terms of some natural 
objects in the $F$-theory. This is important since
the parameters $m_i$ were introduced 
via eq.\refb{ee16}-\refb{ee18} in an {\it ad hoc} 
fashion in order to parametrize the $F$-theory background
and hence one would like to know if there is any reason
why the masses of BPS states should have simple expression
in terms of these parameters. 
We shall now see that these masses can in fact
be expressed in terms of natural objects in the $F$-theory.

For a given open string BPS state starting at the $i$th
seven brane and ending at the $j$-th seven brane, let
us introduced a closed curve $C$ in the $z$ plane that
travels around the $i$th seven brane in anti-clockwise
direction, goes to the $j$th seven brane following the
contour of the open string, travels it in the 
clockwise direction, and comes back to the $i$th
seven-brane by following the contour of the open string
in the opposite direction. Then the mass of the open string
state can be written as
\be \label{ebps1}
\ointop_C \p_z a_D\, ,
\ee
with $a_D$ as defined in ref.\cite{SEIWITTWO}. (Note that we
have renamed the variable $u$ in ref.\cite{SEIWITTWO} as
$z$). To test the validity of eq.\refb{ebps1} we simply
use the fact\cite{SEIWITTWO} that as we move 
around the $i$th seven brane in the anti-clockwise
direction 
\be \label{eada}
\pmatrix{a_D\cr a}\to \pmatrix{a_D+a+m_i\cr a}\, .
\ee
With the help of this equation, \refb{ebps1} reproduces
the mass formula ($m_i-m_j$) for the open string state 
stretched between the $i$th and the $j$th string along the
direct route.
On the other hand, along a closed curve that winds once
around the fifth and the sixth seven branes, the 
vector $\pmatrix{a_D \cr a}$ suffers a monodromy represented
by the matrix\cite{SEIWITTWO}
\be \label{ebps2}
\pmatrix{-1 & 4 \cr 0 & -1}\, .
\ee
Thus the shift $a_D\to a_D-a-m_j$, when transported along
such a curve, corresponds to a shift $a_D\to a_D-a+m_j$. 
As a result, \refb{ebps1} evaluated along the curve $C$ 
associated with the indirect route between the $i$th and the
$j$th seven brane gives an answer $m_i+m_j$. This
agrees with the mass formula obtained from the
orientifold theory. 

We shall now try to reexpress \refb{ebps1} in terms of
integral of the holomorphic two form on the surface
\refb{ee16} along a closed two cycle. For this we note
that\cite{SEIWITTWO}
\be \label{ebps3}
a_D=\ointop_b \lambda\, ,
\ee
where $\lambda$ is the one form introduced in section 17 of
ref.\cite{SEIWITTWO}, and $b$ denotes the $b$ cycle of the
torus represented by eq.\refb{ee16} for fixed $z$.
We can then reexpress \refb{ebps1} as
\be \label{ebps4}
\ointop_C\p_z \ointop_b \lambda
= \ointop_S \omega\, ,
\ee
where 
\be \label{ebps5}
\omega\equiv d\lambda\propto {dx\wedge dz\over y}\, ,
\ee
is the holomorphic two form
on the surface described by eq.\refb{ee16} and $S$ is the
two (real) dimensional surface swept out by the
$b$ cycle of the torus as we move along the closed curve
$C$ in the $z$ plane. Since $a_D$ comes back to $a_D$ up to a
constant shift on being
transported around the curve $C$, the $b$-cycle
of the torus comes back to the $b$ cycle on being transported
around $C$, and hence $S$ is a closed surface. Note that in
defining the surface $S$ we had to choose a specific cycle
on the torus which we defined as the $b$-cycle, and hence
broke manifest $SL(2,Z)$ symmetry of the mass formula. But
this is expected since we are analyzing states that arise
from {\it elementary strings} stretched between two
seven-branes, and elementary strings are not invariant under
SL(2,Z) transformation\cite{SCHW}.
In particular, the mass formula given in eq.\refb{ebps4}
is to be multiplied by the square root of the
string tension of the elementary string.

Eq.\refb{ebps4} gives an expression for the masses of BPS states
in $F$-theory on a (complex) surface in terms of period integrals
of the holomorphic two form on this surface. When one or more of
these period integrals vanish, the surface becomes singular, 
and at the same time the corresponding BPS states become massless, 
signalling the appearance of enhanced gauge symmetries in the
theory.

\noindent{\bf Acknowledgement}:
I wish to thank J. Blum, R. Gopakumar, A. Hanany,
K. Intrilligator, S. Mukhi, C. Vafa and
especially E. Witten for many useful discussions. I would
also like to acknowledge the hospitality of the Institute for
Advanced Study, Princeton, during the course of the work. This
work was partially supported by a grant from Monell Foundation.


\begin{thebibliography}{99}

\bibitem{VAFA}
C. Vafa, hep-th/9602022.

\bibitem{VMOR}
D. Morrison and C. Vafa, hep-th/9602114.

\bibitem{VMORT}
D. Morrison and C. Vafa, hep-th/9603161.

\bibitem{SEIWIT}
N. Seiberg and E. Witten, hep-th/9603003.

\bibitem{WITO}
E. Witten, hep-th/9603150; hep-th/9604030.

\bibitem{FERR}
S. Ferrara, R. Minasian and A. Sagnotti, hep-th/9604097.

\bibitem{ASGR}
P. Aspinwall and M. Gross, hep-th/9605131.

\bibitem{SONE}
A. Sen, hep-th/9603113; hep-th/9604070.

\bibitem{SAGN}
A. Sagnotti, `Open Strings and their Symmetry Groups', Talk at
Cargese Summer Inst., 1987; \\
G. Pradisi and A. Sagnotti, Phys. Lett. {\bf B216} (1989) 59; \\
M. Bianchi, G. Pradisi and A. Sagnotti, Nucl. Phys. {\bf B376}
(1992) 365.

\bibitem{HOR}
P. Horava, Nucl. Phys. {\bf B327} (1989) 461,
Phys. Lett. {\bf B231} (1989) 251.

\bibitem{ORIENT}
E. Gimon and J. Polchinski, hep-th/9601038; \\
J. Polchinski, S. Choudhuri and  C. Johnson, hep-th/9602052; \\
and references therein.

\bibitem{WITTEN}
E. Witten, Nucl. Phys. {\bf B443} (1995) 85 [hep-th/9503124].

\bibitem{DABH}
A. Dabholkar, Phys. Lett.{\bf B357} (1995) 307 [hep-th/9506160].

\bibitem{HULL}
C. Hull, Phys. Lett. {\bf B357} (1995) 545 [hep-th/9506194].

\bibitem{POLWIT}
J. Polchinski and E. Witten, hep-th/9510169.

\bibitem{SEIWITTWO}
N. Seiberg and E. Witten, Nucl. Phys. {\bf B431} (1994) 484
[hep-th/9408099].

\bibitem{COSMIC}
B. Greene, A. Shapere, C. Vafa and S.-T. Yau,
Nucl. Phys. {\bf B337} (1990) 1.

\bibitem{DPTWO}
A. Dabholkar and J. Park, hep-th/9604178.

\bibitem{WISU}
E. Witten, Nucl. Phys. {\bf B460} (1996) 335 [hep-th/9510135].

\bibitem{HETEROTIC}
D. Gross, J. Harvey, E. Martinec and R. Rohm, 
Phys. Rev. Lett.{\bf 54} (1985) 502;
Nucl. Phys. {\bf B256} (1985) 253;
Nucl. Phys. {\bf B267} (1986) 75.

\bibitem{SCHW}
J. Schwarz, Phys. Lett. {\bf B360} (1995) 13 [hep-th/9508143].

\end{thebibliography}
\end{document}